\documentclass[12pt]{article}
\pdfoutput =1
\usepackage{graphics}
\usepackage{graphicx}
\DeclareGraphicsExtensions{.pdf}
\usepackage{float} %Include figure filesusepackage{graphicx} %Include figure files
\usepackage{subfloat}
\usepackage{amsmath}
\usepackage{slashed}
\usepackage{amsfonts}  
\usepackage{amssymb}

\def\g{\gamma}
\def\d{\delta}

\def\S{\Sigma}

\usepackage{color}

\usepackage{subcaption}
\begin{document}
\date{}
%%%%%%%%%%%%%%%%%%%%
\title{{\bf{\Large Holographic central charge for sine dilaton gravity}}}
%%%%%%%%%%%%%%%%%%%%
\author{
{\bf { ~Paramesh Mahapatra}$
$\thanks{E-mail:  paramesh\_m@ph.iitr.ac.in}}\\
 {\normalsize  Department of Physics, Indian Institute of Technology Roorkee,}\\
  {\normalsize Roorkee 247667, Uttarakhand, India}
\\[0.3cm]
{\bf { Hemant Rathi}$
$\thanks{E-mail:  hemant.rathi@saha.ac.in}}\\
 {\normalsize  Saha Institute of Nuclear Physics, 1/AF Bidhannagar, }\\
  {\normalsize Kolkata 700064, India }
  \\[0.3cm]
 {\bf { ~Dibakar Roychowdhury}$
$\thanks{E-mail:  dibakar.roychowdhury@ph.iitr.ac.in}}\\
 {\normalsize  Department of Physics, Indian Institute of Technology Roorkee,}\\
  {\normalsize Roorkee 247667, Uttarakhand, India}
\\[0.3cm]
}
\maketitle
%%%%%%%%%%%%%%%%%%%%%%%%%%%%%%%%%%%%%%%%%%%%%%%%%%%%%%%%%%%%
\abstract{ In this paper, we calculate the holographic central charge ($c_{sDG}$) associated with sine dilaton gravity (sDG) in the semiclassical limit. In the low energy limit, the sDG flows into the ordinary JT gravity which is a conjectured dual of ordinary Schwarzian quantum mechanics at strong coupling. We compute the associated central charge ($c_{JT}$), which reveals $c_{sDG}>c_{JT}$. We identify this as an artifact of the UV completion of JT gravity in terms of sine dilaton gravity.}
%%%%%%%%%%%%%%%%%%%%%%%%%%%%%%%%%%%%
\section{Overview and motivation }
 One of the most simplest and intriguing realizations of holography \cite{Maldacena:1997re}-\cite{Gubser:1998bc} in the present era has been the correspondence between the Jackiw-Teitelboim gravity (JT) in (1+1) dimensions \cite{Jackiw:1984je}-\cite{Teitelboim:1983ux} and the Sachdev-Ye-Kitaev (SYK) model in (0+1) dimension \cite{Sachdev:1992fk}-\cite{Gross:2016kjj}. The SYK model \cite{Sachdev:1992fk}-\cite{Maldacena:2016hyu} has drawn renewed attention in the recent years due its several remarkable features. For example, it is a solvable model of quantum many-body interactions at strong coupling and in the Large $N$ limit. On the other hand, its bulk dual, namely the JT gravity, has been proved to be an excellent toy model of quantum gravity in two dimensions \cite{Mertens:2022irh}. 
 
 Interestingly enough, the above duality that holds in the deep IR of a RG-flow, has recently been realized to be uplifted in the sense of a UV completion of a (0+1)d QFT \cite{Blommaert:2023opb}-\cite{Blommaert:2024whf}. The above correspondence goes under name of the conjectured duality between the double scaled SYK (DSSYK) (which is regarded as the all energy generalization of SYK) \cite{Blommaert:2023opb}-\cite{Berkooz:2024lgq} and the sine dilaton gravity (sDG) \cite{Blommaert:2024ydx}-\cite{Blommaert:2024whf}, which has been claimed to be the UV completion of JT gravity. 

   In this paper, we compute the holographic central charge \cite{Brown:1986nw}-\cite{Laflorencie:2015eck} associated with the sDG ($c_{sDG}$) and in its low-energy regime, namely as the JT gravity ($c_{JT}$), using the variational properties of the boundary stress-energy tensor under the combined action of diffeomorphism and U(1) gauge transformation. Our analysis provides substantial evidence for sDG to be the UV completion theory of JT gravity, where we show $c_{sDG}>c_{JT}$ .

    The rest of the paper is organized as follows:
    
    $\bullet$ In Section \ref{sdgbody}, we study the sDG in the presence of $U(1)$ gauge interaction. In particular, we obtain solutions for background fields in the Fefferman Graham (FG) gauge and compute boundary stress-energy tensor ($T_{tt}$). Additionally, we investigate the transformation properties of $T_{tt}$ under the combined action of diffeomorphism and $U(1)$ gauge transformation and obtain the holographic central charge. 

    $\bullet$ In Section \ref{jtbody}, we explore the low energy limit of sDG keeping the $U(1)$ gauge potential. This turns out to be the well-known JT gravity coupled with $U(1)$ gauge fields. We compute the associated central charge using holographic techniques and compare it to that with the sDG.

    $\bullet$ Finally in Section \ref{cncbody}, we conclude our discussion along with mentioning some interesting future extensions.

\section{Periodic U(1) dilaton gravity }\label{sdgbody}
The complete action for sDG \cite{Blommaert:2024ydx} in the presence of U(1) gauge potential ($A_{\mu}$) can be expressed as 
\begin{equation}\label{action}
    I_{sDG} = \frac{1}{4|\log q|}\int d^2x \sqrt{-g}[\Phi R + 2 \sin[\Phi] - \frac{\xi}{4}F_{\mu\nu}F^{\mu\nu}]+I_{bdy},
\end{equation}
where $F_{\mu\nu}$ is the electromagnetic field strength tensor for the U(1) gauge field ($A_{\mu}$) and $\xi$ is the associated (dimensionless) coupling constant. 

Here, $I_{bdy}$ stands for the boundary action, which gets modified in the presence of U(1) potential \cite{Rathi:2023vhw}
\begin{align}\label{bdyact}
    I_{bdy}=\frac{1}{2|\log q|}\int dt \sqrt{-\g}[\Phi K -ie^{-i\Phi/2} + \frac{\xi}{4}\sqrt{\g^{ab}A_aA_b}]\hspace{1mm},\hspace{1mm}|\log q|=\frac{p^2}{N}
\end{align}
where $(a,b)$ are the boundary indices.  Notice that, the action (\ref{bdyact}) is free from boundary divergences where $\g^{ab}$ is the induced boundary metric. In the semiclassical limit, $|\log q|<<1$.

\subsection{Dynamics of background fields}
On varying the bulk action (\ref{action}) with respect to the background fields, we obtain the following set of equations 
\begin{align}
     \left(\bigtriangledown_\mu \bigtriangledown_\nu - g_{\mu\nu}\square\right)\Phi + \frac{\xi}{2}\left(F_{\mu\rho}F_\nu^\rho - \frac{1}{4}F^2g_{\mu\nu}\right) + \sin(\Phi) g_{\mu\nu} = 0,\label{metric} \\
  R + 2\cos(\Phi)= 0, \\
   \bigtriangledown_\mu F^{\mu\nu}=0.\label{gauge}
\end{align}

Next, we solve these equations (\ref{metric})-(\ref{gauge}) in the Fefferman Graham (FG) gauge \cite{Castro:2008ms}, \cite{fefferman}. In this gauge, the computation of holographic central charge \cite{Castro:2008ms} turns out to be relatively easy. The space-time metric and gauge field ansatz in the FG gauge can be expressed as  
\begin{align}
    ds^2 = d\eta^2 + h_{tt}(\eta)dt^2, \label{metricfg}\\
 A_\mu dx^\mu = A_t(\eta) dt\hspace{1mm}, \hspace{1mm} \Phi\equiv\Phi(\eta).\label{gfg}
\end{align}

In FG gauge (\ref{metricfg})-(\ref{gfg}), the above equations (\ref{metric})-(\ref{gauge}) boil down into 
\begin{align}
    \frac{h'_{tt}(\eta)^2 - 2h_{tt}(\eta)h''_{tt}(\eta)}{2 h_{tt}(\eta)^2} + 2 \cos\left(\Phi(\eta)\right) = 0, \label{metriceqn}\\
    \Phi''(\eta) - \frac{h'_{tt}(\eta)}{2h_{tt}(\eta)} \Phi'(\eta) = 0, \\ 
    \partial_\eta\left(\frac{A_t'(\eta)}{\sqrt{-h_{tt}(\eta)}}\right) = 0, \label{gaugeeqn}
\end{align}
where $'$ denotes the derivative with respect to the variable $\eta$. Notice that, the above sets of equations (\ref{metriceqn})-(\ref{gaugeeqn}) are quite difficult to solve due to the presence of the cosine function in (\ref{metriceqn}). However, one can solve these equations (\ref{metric})-(\ref{gauge}) in a new gauge, where the metric ansatz takes the following form \cite{Blommaert:2024ydx}
 \begin{align}
      ds^2 = F(r) dt^2 + \frac{dr^2}{F(r)}. \label{newgauge}
  \end{align}
  
In the above gauge (\ref{newgauge}), the equations (\ref{metric})-(\ref{gauge}) can be expressed as 
\begin{align}
     \Phi''(r) = 0, \label{metric2}\\
      F''(r) - 2 \cos(\Phi(r)) = 0,\\
      A_t''(r) = 0,\label{gauge2}
\end{align}
where $'$ denotes derivative with respect to the radial variable $r$.

The solutions to the above set of equations (\ref{metric2})-(\ref{gauge2}) are given below
\begin{align}
     \Phi(r) = &\hspace{1mm}c_1 r + c_2,\label{phir}\\
      F(r) = &\hspace{1mm}c_5 +c_6 r - \frac{2 \cos(c_1 r+c_2 )}{c_1^2},\label{Fr}\\
      A_t(r) = &\hspace{1mm}c_3 r + c_4,\label{Atr}
\end{align}
where $c_i$, $(i=1,2,..6)$ are the integration constants.

Notice that, in this gauge, the space-time boundary is located\footnote{Notice that, $g_{tt}$ diverges at the boundary. In particular, $\cos r$ diverges at $r=\pi/2+i\infty.$ } at $r=\frac{\pi}{2}+\text{i}\infty$ and the following boundary conditions hold \cite{Blommaert:2024ydx}
\begin{align}\label{bcb}
\sqrt{F}e^{i\Phi_{bdy}/2}=i\hspace{1mm},\hspace{3mm}\Phi_{bdy}=\frac{\pi}{2}+\text{i}\infty.
\end{align}

It is interesting to notice that one can switch back to our previous FG gauge (\ref{metricfg})-(\ref{gfg}), using the following coordinate redefinition \cite{Rathi:2021aaw}
\begin{align}\label{trans}
    \eta=\int \frac{dr}{\sqrt{F(r)}},
\end{align}
and identify $h_{tt}(\eta)=F(r)$, where $r$ should be expressed in terms of the variable $\eta$.

Using (\ref{Fr}) and (\ref{trans}), one finds the precise mapping between $r$ and FG variable $\eta$ as\footnote{Here, we set $c_1=c_3=1,\hspace{1mm}c_2=c_4=c_6=0$ and $c_5=3$. }
\begin{align}
      r =  2 \text{JacobiAmplitude} \Big[\frac{\eta}{2}, -4 \Big].\label{r}
\end{align}
It is noteworthy to mention that the equations of motion in FG gauge\footnote{In the $\eta$ variable, the boundary is located at $\eta_{bdy}=2.01+1.48i$, such that $h_{tt}(\eta)$ diverges.} (\ref{metriceqn})-(\ref{gaugeeqn}) is also satisfied subjected to the coordinate transformation (\ref{r}).

Using (\ref{r}), one can express the background fields in FG coordinate
\begin{align}
    h_{tt}(\eta)=& \hspace{1mm}-2\cos\left(  2 \text{JacobiAmplitude} \Big[\frac{\eta}{2}, -4 \Big]\right)+3,\label{fgsol1}\\
    \Phi(\eta)=&  \hspace{1mm} 2 \text{JacobiAmplitude} \Big[\frac{\eta}{2}, -4 \Big],\\
    A_t(\eta)=& \hspace{1mm}  2 \text{JacobiAmplitude} \Big[\frac{\eta}{2}, -4\Big],\label{fgsol3}
\end{align}
which when substituted back into (\ref{metriceqn})-(\ref{gaugeeqn}) yields a trivial solution\footnote{See Appendix \ref{Aa}, for a detailed numerical analysis.} .

\subsection{Holographic central charge}
In this Section, we first obtain the holographic stress-energy tensor \cite{Castro:2008ms}-\cite{Rathi:2021aaw} associated with the 1D boundary theory and subsequently investigate its transformation properties under the mutual effects of diffeomorphism and $U(1)$ gauge transformation. Finally, we compute the holographic central charge using these properties. 

The boundary stress-energy tensor \cite{Castro:2008ms}-\cite{Rathi:2021aaw} is defined as 
\begin{align}\label{stt}
    T^{ab} = \frac{2}{\sqrt{-\gamma}} \frac{\delta I_{bdy}}{\delta \gamma^{ab}}.
\end{align}

Using (\ref{bdyact}) and (\ref{stt}), we obtain
\begin{align}
    T^{ab}=&\hspace{1mm}\frac{1}{2|\log q|}\Bigg[n_\mu \bigtriangledown^\mu \Phi \gamma^{ab} + n^\mu \frac{\Phi}{\sqrt{-\gamma}} \left(\partial_\mu \sqrt{-\gamma} \right)\gamma^{ab} - i \gamma^{ab} e^{-i \frac{\Phi}{2}} + \frac{\xi}{4} \gamma^{ab}\left(\sqrt{\gamma^{cd} A_c A_d}\right) \nonumber\\
  &\hspace{1mm} + \frac{\xi}{4}\left(\frac{A^a A^b}{\sqrt{\gamma^{cd} A_c A_d}}\right) \Bigg],\label{stresstensor}
\end{align}
where $n^{\mu}=\delta^{\mu}_{\eta}$ denotes the unit normal vector \cite{Rathi:2023vhw}. 

In the FG gauge (\ref{metricfg})-(\ref{gfg}), the above stress-energy tensor (\ref{stresstensor}) boils down into  
\begin{align}\label{sDGexpst}
    T_{tt}=\frac{1}{2|\log q|}\left(h_{tt}\partial_{\eta}\Phi+\frac{\Phi}{2}\partial_{\eta}h_{tt}-h^{3/2}_{tt}+\frac{\xi}{2}A_t\sqrt{h_{tt}}\right),
\end{align}
where we have used the boundary condition (\ref{bcb}).

Next, we study the transformation properties of (\ref{stresstensor}) under the diffeomorphism: $x^{\mu}\rightarrow x^{\mu}+\epsilon^{\mu}(x^{\alpha})$, where $\epsilon^{\mu}(x^{\alpha})$ are the parameters associated with the diffeomorphism.

Under diffeomorphism, the background fields transform as 
\begin{align}
    \delta_\epsilon A_\mu = &\hspace{1mm} \epsilon^\nu\bigtriangledown_\nu A_\mu + A_\nu\bigtriangledown_\mu\epsilon^\nu, \label{gaugetransform}\\
      \delta_\epsilon g_{\mu\nu} = &\hspace{1mm} \bigtriangledown_\mu\epsilon_\nu + \bigtriangledown_\nu \epsilon_\mu, \label{metrictransfrom}\\
      \delta_\epsilon \Phi = &\hspace{1mm} \epsilon^\mu\bigtriangledown_\mu\Phi. \label{phitransform}
\end{align}

One can read off diffeomorphism parameters using (\ref{gaugetransform})-(\ref{phitransform}). Using background solutions (\ref{phir})-(\ref{Atr}) along with the map (\ref{r}), one can show
\begin{align}\label{deff}
   \epsilon_t= G(t) \left[2 \cos \left(2\text{JacobiAmplitude}\left[\frac{\eta}{2},4\right]\right)-3\right]\hspace{1mm},\hspace{2mm}\epsilon_{\eta}=F,
\end{align}
where $F$ and $G(t)$ are the integration constants. 

It is important to notice that we are working in a gauge where $A_{\eta}=0$ (\ref{gfg}). Therefore any coordinate transformation must respect this condition. To verify this, we consider the variation of $A_{\eta}$ under diffeomorphism, which yields
\begin{align}\label{sggp}
    \delta_\epsilon A_\eta = A_t \partial_\eta\epsilon^t\hspace{1mm},\hspace{1mm}\epsilon^t= - G(t).
\end{align}

Substituting (\ref{deff}) into (\ref{sggp}) and after performing differentiation, one finds that variation $\delta_{\epsilon}A_{\eta}$ vanishes identically. Therefore, the above diffeomorphism (\ref{deff}) is consistent as it preserves the gauge condition (\ref{gfg}). 

Next, we express the complete variation of the boundary stress-energy tensor (\ref{sDGexpst}) under diffeomorphism,
\begin{align}\label{varstsdg}
    \delta_{\epsilon}T_{tt}=&\hspace{1mm}\frac{1}{2|\log q|}\Bigg[(\delta_{\epsilon} h_{tt})\partial_{\eta}\Phi+h_{tt}\partial_{\eta}(\delta_{\epsilon} \Phi)+\frac{(\delta_{\epsilon} \Phi)}{2}\partial_{\eta}h_{tt}+\frac{\Phi}{2}\partial_{\eta}(\delta_{\epsilon}  h_{tt})-\frac{3}{2}\sqrt{h_{tt}}\delta_{\epsilon} h_{tt}\nonumber\\
    &\hspace{1mm}\frac{\xi}{2}(\delta_{\epsilon}  A_t)\sqrt{h_{tt}}+\frac{\xi}{2}A_t\delta_{\epsilon} (\sqrt{h_{tt}})\Bigg],
\end{align}
where the variation of background fields (\ref{gaugetransform})-(\ref{phitransform}) can be expressed as
\begin{align}
    \delta_\epsilon A_t = &\hspace{1mm} F \left(\text{JDN}\Big[\frac{\eta}{2},-4\Big]\right)  + \partial_t G(t) \left( -2 \text{JA}\Big[\frac{\eta}{2},-4\Big]\right)\label{exp1},\\
    \delta_\epsilon h_{tt} = &\hspace{1mm}2F \left( \text{JDN}\Big[\frac{\eta}{2},-4\Big] \sin\Big[2 \text{JA}\Big[\frac{\eta}{2},-4\Big]\Big]\right) + 2\partial_t G(t) \left( -3 + 2 \cos\left(2 \text{JA}\Big[\frac{\eta}{2},-4\Big]\right)\right),\\
       \delta_\epsilon \Phi = &\hspace{1mm}F \left(\text{JDN}\Big[\frac{\eta}{2},-4\Big]\right)\label{exp3},
\end{align}
where JA and JDN respectively denote the JacobiAmplitude and JacobiDN \cite{formula}.

Using (\ref{stresstensor}), (\ref{varstsdg})-(\ref{exp3}), one can finally express the variation of $T_{tt}$ as
\begin{align}
    \delta_{\epsilon} T_{tt} = 2 T_{tt}F - \frac{c_{sDG}}{24 \pi} \partial_tG(t),
\end{align}
where, we identify the central charge of sDG as\footnote{In the above derivation, we replace FG variable $\eta$ using the inverse map of (\ref{r}) and replace $r$ by $r_{bdy}=\pi/2+i \infty$. Here, we absorb the imaginary contribution through the redefinition of constants, $F$ and $G$.} 
\begin{align}\label{SDGCC}
    c_{sDG}=\frac{24\pi}{25|\log q|}\left(3+34\xi\right).
\end{align}
Notice that, the $c_{sDG}$ grows as $\sim\frac{1}{|\log q|}$ and $|\log q|<<1$. Therefore it is a large number in the semiclassical limit. Moreover, the central charge (\ref{SDGCC}) boil down to ordinary sDG in the absence of the U(1) gauge field ($\xi= 0$).

\section{Comments on JT gravity limit} \label{jtbody}
We now consider the low energy limit of periodic U(1) dilaton gravity (\ref{action}). This is achieved by taking the limit $|\log q|\rightarrow0$ \cite{Blommaert:2024whf}, which eventually leads to the well-known JT gravity in the presence of a gauge field \cite{Castro:2008ms}. We derive the associated boundary stress-energy tensor and compute the holographic central charge following the standard procedure as discussed in Section 2.

To begin with, we rescale the dilaton ($\Phi$) and the coupling constant ($\xi$) as 
\begin{align}\label{scaling}
    \Phi=2\pi+2|\log q|\Phi_{JT}\hspace{1mm},\hspace{2mm}\xi\rightarrow2\xi|\log q|,
\end{align}
where $\Phi_{JT}$ is the dilaton for the JT gravity \cite{Castro:2008ms}. 

Next, we take the low energy limit $|\log q|\rightarrow0$ keeping $\Phi_{JT}$ and $\xi$ finite, which results in a JT gravity action \cite{Castro:2008ms}
\begin{equation}\label{jtaction}
    I_{JT} = \frac{1}{16\pi G_2}\int d^2x \sqrt{-g}[ \Phi_{JT}(R + 2) - \frac{\xi}{4}F_{\mu\nu}F^{\mu\nu}]\hspace{1mm}+I_{bdy},
\end{equation}
where we denote the GHY term as\footnote{In order to be consistent with the literature \cite{Castro:2008ms}, we include a factor of $1/16\pi G_2$ in front of the action by hand, where $G_2\sim\frac{1}{N}$ is the gravitational coupling constant in JT gravity. For the rest of the paper, we drop the subscript ($JT$) in dilaton ($\Phi$). }. 
\begin{align}\label{jtbdyact}
    I_{bdy}=\frac{1}{8\pi G_2}\int dt \sqrt{-\g}[\Phi_{JT} K +\frac{\Phi_{JT} }{2} + \frac{\xi}{4}\sqrt{\g^{ab}A_aA_b}].
\end{align}

\subsection{Dynamics of background fields}

The equations of motion associated with the background fields ($\Phi$, $g_{\mu\nu}$ and $A_{\mu}$) can be obtained by varying the action (\ref{jtaction})
\begin{align}
     \left(\bigtriangledown_\mu \bigtriangledown_\nu - g_{\mu\nu}\square\right)\Phi + \frac{\xi}{2}\left(F_{\mu\rho}F_\nu^\rho - \frac{1}{4}F^2g_{\mu\nu}\right) + \Phi g_{\mu\nu} = 0,\label{jtmetric} \\
  R + 2= 0, \\
   \bigtriangledown_\mu F^{\mu\nu}=0.\label{jtgauge}
\end{align}

Next, we solve these dynamical equations (\ref{jtmetric})-(\ref{jtgauge}) in the static FG gauge (\ref{metricfg})-(\ref{gfg}), which yields 
\begin{align}
    \partial_{\eta}^2\sqrt{-h_{tt}(\eta)}-\sqrt{-h_{tt}(\eta)} = 0, \label{jtmetriceqn}\\
    \Phi''(\eta) - \frac{h'_{tt}(\eta)}{2h_{tt}(\eta)} \Phi'(\eta) = 0, \\ 
    \partial_\eta\left(\frac{A_t'(\eta)}{\sqrt{-h_{tt}(\eta)}}\right) = 0.\label{jtgaugeeqn}
\end{align}

Upon solving these equations (\ref{jtmetriceqn})-(\ref{jtgaugeeqn}), one finds
\begin{align}
    \sqrt{-h_{tt}(\eta)}=&\hspace{1mm} m_1e^{\eta}+ m_2e^{-\eta},\label{jtsol1}\\
    A_t(\eta)=&\hspace{1mm} m_3(m_1e^{\eta}+ m_2e^{-\eta})+m_4,\label{jtat}\\
    \Phi(\eta)=&\hspace{1mm} m_5(m_1e^{\eta}+ m_2e^{-\eta})+m_6,\label{jtsol2}
\end{align}
where $m_i$, $(i=1,2..6)$ are integration constants. 

\subsection{Stress-energy tensor and central charge}
We now compute the boundary stress-energy tensor and study its properties under the combined action of diffeomorphism and U(1) gauge transformations. Finally, we obtain the central charge for the boundary theory and compare it with the central charge for sDG gravity obtained previously.

The stress-energy tensor (\ref{stt}) can be obtained by varying the action (\ref{jtbdyact}) with respect to the induced metric ($\g^{ab}$) 
\begin{align}
    T^{ab}=&\hspace{1mm}\frac{1}{8\pi G_2}\Bigg[n_\mu \bigtriangledown^\mu \Phi \gamma^{ab} + n^\mu \frac{\Phi}{\sqrt{-\gamma}} \left(\partial_\mu \sqrt{-\gamma} \right)\gamma^{ab} + \gamma^{ab} \frac{\Phi}{2} + \frac{\xi}{4} \gamma^{ab}\left(\sqrt{\gamma^{cd} A_c A_d}\right) \nonumber\\
  &\hspace{1mm} + \frac{\xi}{4}\left(\frac{A^a A^b}{\sqrt{\gamma^{cd} A_c A_d}}\right) \Bigg].\label{jtstresstensor}
\end{align}

In the FG gauge (\ref{metricfg})-(\ref{gfg}), above entity (\ref{jtstresstensor}) boils down into  
\begin{align}\label{jtexpst}
    T_{tt}=\frac{1}{8\pi G_2}\left(h_{tt}\partial_{\eta}\Phi+\frac{\Phi}{2}\partial_{\eta}h_{tt}+\frac{1}{2}h_{tt}\Phi+\frac{\xi}{2}A_t\sqrt{h_{tt}}\right).
\end{align}

Next, we explore the transformation properties of (\ref{jtexpst}) under the combined action of diffeomorphism and U(1) gauge transformations. To begin with, we compute the diffeomorphism parameter ($\epsilon^{\mu}(x^{\alpha})$) using the variations of background fields (\ref{gaugetransform})-(\ref{phitransform}) and the solutions (\ref{jtsol1})-(\ref{jtsol2}), which yields
\begin{align}\label{jtdef}
    \epsilon_{\eta}=f(t)\hspace{1mm},\hspace{2mm}
     \epsilon_{t}=\frac{e^{-2\eta}}{2m_1}\left(m_2+m_1e^{2\eta}\right)\left(2m_1\left(m_2+m_1e^{2\eta}\right)g(t)+\partial_tf(t)\right),
\end{align}
where $f(t)$ and $g(t)$ are the integration constants.

Notice that under diffeomorphism (\ref{jtdef}), one of the components of U(1) gauge field transforms as 
\begin{align}
    \delta_\epsilon A_\eta = A_t \partial_\eta\epsilon^t\neq0,
\end{align}
which violates the gauge condition (\ref{gfg}), where $A_{\eta}$ is set to be zero. In order to restore the gauge condition, we employ a second U(1) gauge transformation, $A_{\mu}\rightarrow A_{\mu}+\partial_{\mu}\S$ and choose $\S$ such that $\d_{(\epsilon+\S)}A_{\eta}=0.$

Using (\ref{jtat}) and (\ref{jtdef}), we obtain
\begin{align}\label{jtsigma}
    \S=-\int d\eta A_t\partial_{\eta}\left(\frac{\epsilon_t}{h_{tt}}\right)=\frac{\partial_tf(t)\left(m_4+2m_1m_3e^{\eta}\right)}{2m_1\left(m_2+m_1e^{2\eta}\right)}.
\end{align}

Below, we note down the variation of the boundary stress energy tensor (\ref{jtexpst}) under the combined action of diffeomorphism and U(1) gauge transformation, which yields
\begin{align}\label{varst}
    \delta_{(\epsilon+\S)}T_{tt}=&\hspace{1mm}\frac{1}{8\pi G_2}\Bigg[(\delta_{\epsilon} h_{tt})\partial_{\eta}\Phi+h_{tt}\partial_{\eta}(\delta_{\epsilon} \Phi)+\frac{(\delta_{\epsilon} \Phi)}{2}\partial_{\eta}h_{tt}+\frac{\Phi}{2}\partial_{\eta}(\delta_{\epsilon}  h_{tt})+ \frac{1}{2}(\delta_{\epsilon}  h_{tt})\Phi+\nonumber\\
    &\hspace{1mm}\frac{1}{2}h_{tt}(\delta_{\epsilon}  \Phi)+\frac{\xi}{2}(\delta_{(\epsilon+\S)}  A_t)\sqrt{h_{tt}}+\frac{\xi}{2}A_t\delta_{\epsilon} (\sqrt{h_{tt}})\Bigg],
\end{align}
where the individual variations of background fields (\ref{gaugetransform})-(\ref{phitransform}) can be summarised as
\begin{align}
    \delta_{(\epsilon+\S)}A_t=&\frac{1}{2 m_1 \left(m_1e^{2 \eta }+m_2\right)}\Bigg[e^{-\eta }\Big(\left(m_2m_3-e^{\eta } \left(m_1m_3 e^{\eta }+m_4\right)\right) \big(\partial_{t}^2f(t)+2m_1\partial_tg(t)\times\nonumber\\
    &\left(m_1 e^{2 \eta }+m_2\right)\big)+2 m_1 m_3 f(t) \left(m_1 e^{2 \eta }+m_2\right)^2\Big)+ \partial_t^2f(t)\left(m_4+2m_1m_3e^{\eta}\right)\Bigg],\label{fieldvar1}\\
    \delta_{\epsilon}h_{tt}=&\hspace{1mm}\frac{e^{-2 \eta } }{m_1}\left(m_1 e^{2 \eta }+m_2\right) \big(\partial_t^2f(t)+2 m_1 \big(f(t) \left(m_2-m_1 e^{2 \eta }\right)+\nonumber\\
    &\hspace{1mm}\partial_tg(t) \left(m_1 e^{2 \eta }+m_2\right)\big)\big),\\
    \delta_{\epsilon}\Phi=&\hspace{1mm} f(t) m_5\left(m_1 e^{\eta }+m_2 e^{-\eta }\right).\label{fieldvar3}
\end{align}

Using (\ref{jtexpst}) and (\ref{fieldvar1})-(\ref{fieldvar3}), one can finally express (\ref{varst}) as  
\begin{align}
   \delta_{(\epsilon+\S)}T_{tt}=2T_{tt}f(t)-\frac{c_{JT}}{24\pi}\partial_{t}^2f(t), 
\end{align}
where we identify
\begin{align}\label{fcjt}
    c_{JT}=\frac{5\xi}{2G_2}
\end{align}
as the holographic central charge of the JT gravity model (\ref{jtaction}).

It is important to notice that the $c_{JT}$ vanishes in the absence of U(1) gauge fields. Similar observation were made earlier by authors in \cite{Castro:2008ms}-\cite{Rathi:2023vhw}. Whereas on the other hand, for sDG we still have a non-zero central charge for $\xi=0$. 

In order make a precise comparison between the two central charges, we now rewrite the $c_{sDG}$ (\ref{SDGCC}) in terms of $c_{JT}$ (\ref{fcjt}), which yields
\begin{align}\label{comp}
   c_{sDG}=\frac{8\pi }{25 |\log q|}\left(9+\frac{204}{5}\frac{|\log q|}{p^2}c_{JT}\right).
\end{align}

The above expression (\ref{comp}) clearly shows that sDG has more degree of freedom than JT gravity. This can be nicely associated with the notion of a flow central charge or $c$-function \cite{Macpherson:2014eza}-\cite{Chatzis:2024kdu} that monotonically decreases in a RG flow such that $c_{sDG}>c_{JT}$. In other words, JT gravity can be interpreted as the deep IR limit of its UV completion namely the sDG.

\section{Conclusion and future direction}\label{cncbody}
To sum up, in the present work, we explore the holographic aspects of sine dilaton gravity (sDG) coupled to U(1) gauge field. In particular, we compute solutions for background fields and obtain the boundary stress-energy tensors in the FG gauge. We examine the variational properties of these background solutions and boundary stress-energy tensor under the diffeomorphism. We find that under diffeomorphism, the variation of the gauge field preserves the gauge condition. We finally compute the holographic central charge ($c_{sDG}$) associated with the sDG using these variational properties of the boundary stress-energy tensor.

Next,  we examine the low-energy limit of sDG keeping the U(1) gauge interaction intact, which boils down into the well-known JT gravity coupled with a U(1) potential. We obtain corresponding solutions for background fields and compute the boundary stress-energy tensor and investigate its properties under diffeomorphism. Unlike the sDG, here, diffeomorphism breaks the gauge condition, and hence, we require an additional U(1) gauge transformation to restore the FG gauge condition. Finally, we compute the holographic central charge ($c_{JT}$) by utilizing the variational properties of the boundary stress-energy tensor under the combined action of diffeomorphism and U(1) gauge transformations. We also compare the two central charges, which shows $c_{sDG}> c_{JT}$.

Below, we outline some interesting future extensions of our present paper. 

$\bullet$ A two site SYK model with complex couplings has been discussed in \cite{Garcia-Garcia:2020ttf}. An interesting observation that emerges from the discussion is that at low energies (without introducing a coupling between the two sites), the free energy becomes constant with the temperature up to a critical point where it changes abruptly, thereby indicating the onset of a first order phase transition in the system. The dual to the constant free energy phase on the JT gravity side is a charged wormhole solution. Further in  \cite{Rathi:2021mla}, the phases of the wormhole solution has been discussed. It has been shown that as the temperature is varied from low to high, the system undergoes a phase transition from a wormhole to two separate black holes. Therefore, it would be an interesting problem to see if a similar setup in sDG produces an analogue wormhole to black hole phase transition. Recently, wormhole amplitude in sDG has been computed in \cite{Blommaert:2025avl} using the WdW quantization. The amplitude has been shown to match with the spectral correlation of finite-cut matrix integral, thereby strengthening the relation of sDG with q-deformed JT gravity matrix integral. \par  
      \par
$\bullet$ It has been shown in literature that there also exists a de-Sitter (dS) version of JT gravity (dS/JT gravity) \cite{Maldacena:2019cbz}-\cite{Held:2024rmg}. The sDG also yields the dS/JT gravity by using an appropriate scaling (redefinition of the dilaton field) as discussed in \cite{Blommaert:2024whf}. The dS/JT gravity is the conjectured dual of DSSYK at high temperature \cite{Rahman:2022jsf}-\cite{Susskind:2022bia}. Therefore, a natural extension of our present work would be to map the central charge of sDG to that into the central charge of dS gravity as found in \cite{Verlinde:2024zrh}.
\par

We hope to address some of these issues in our future work.
\section*{Acknowledgments}
PM and DR are indebted to the authorities of Indian Institute of Technology, Roorkee for their unconditional support towards researches in basic sciences. HR would like to thank the authorities of Saha Institute of Nuclear Physics, Kolkata, for their support. DR would like to acknowledge The Royal Society, UK for financial assistance. DR also acknowledges the Mathematical Research Impact Centric Support (MATRICS) grant (MTR/2023/000005) received from ANRF, India.

\appendix
\section{Numerical solution}\label{Aa}
In this Section, we show that the (\ref{fgsol1})-(\ref{fgsol3}) are indeed the solutions of the equations of motion (\ref{metriceqn})-(\ref{gaugeeqn}) in FG gauge. To start with, we rewrite these equations (\ref{metriceqn})-(\ref{gaugeeqn}) in the following way 
\begin{align}
\partial_{\eta}^2\sqrt{h_{tt}(\eta)}=&\hspace{1mm}\cos \left(\Phi(\eta)\right)\sqrt{h_{tt}(\eta)}, \label{Aa1}\\
    \Phi'(\eta)^2=&\hspace{1mm}h_{tt}(\eta), \label{Aa2}\\
      A_t'(\eta)^2=&\hspace{1mm}h_{tt}(\eta). \label{Aa3}
\end{align}

\begin{figure}[htp]
\begin{center}
\includegraphics[scale=.55]{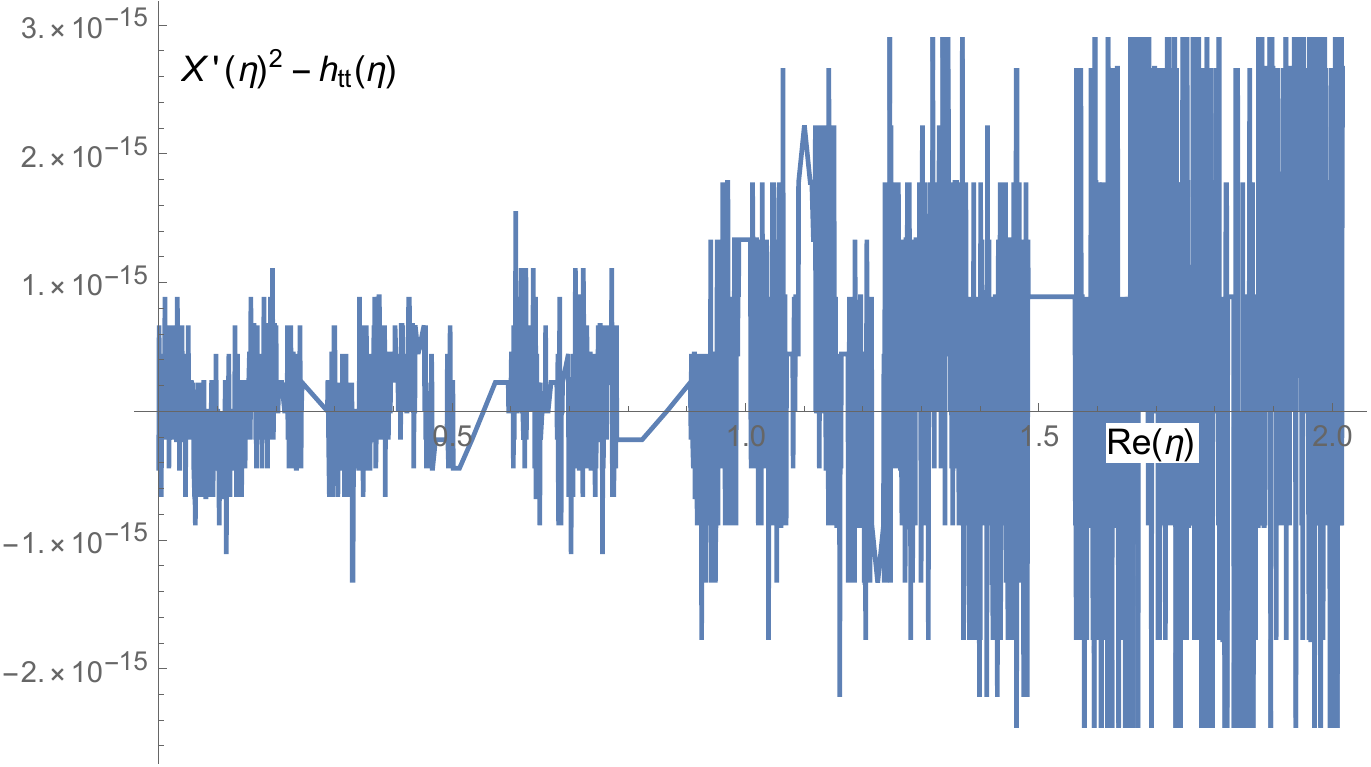}
\caption{Plot of the function $X'(\eta)^2-h_{tt}(\eta)$ vs $\eta$, where $X(\eta)=\left(\Phi(\eta),\hspace{1mm} A_{t}(\eta)\right)$. } 
\label{FIGURE}
\end{center}
\end{figure}

Notice that the equation (\ref{Aa1}) is trivially satisfied by the solutions (\ref{fgsol1})-(\ref{fgsol3}). On the other hand, it is quite challenging to check that the expressions (\ref{fgsol1})-(\ref{fgsol3}) are also the solutions of (\ref{Aa2}) and (\ref{Aa3}). Therefore, we will check them numerically and in particular, we plot the function $X'(\eta)^2-h_{tt}(\eta)$, where $X(\eta)=\left(\Phi(\eta),\hspace{1mm} A_{t}(\eta)\right)$ against the real value of the variable $\eta$ as shown in figure (\ref{FIGURE}).

It is clear from the above figure (\ref{FIGURE}) that the difference between the left-hand side and right-hand side of the equations (\ref{Aa2})-(\ref{Aa3}) is of the order $10^{-15}$, which can be ignored for the practical purpose. In other words, the solutions (\ref{fgsol1})-(\ref{fgsol3}) satisfy the equations of motion (\ref{Aa1})-(\ref{Aa3}).


\begin{thebibliography}{99}

\bibitem{Maldacena:1997re}
J.~M.~Maldacena,
``The Large N limit of superconformal field theories and supergravity,''
Adv. Theor. Math. Phys. \textbf{2} (1998), 231-252
doi:10.1023/A:1026654312961
[arXiv:hep-th/9711200 [hep-th]].

\bibitem{Witten:1998qj}
E.~Witten,
``Anti-de Sitter space and holography,''
Adv. Theor. Math. Phys. \textbf{2} (1998), 253-291
doi:10.4310/ATMP.1998.v2.n2.a2
[arXiv:hep-th/9802150 [hep-th]].

\bibitem{Gubser:1998bc}
S.~S.~Gubser, I.~R.~Klebanov and A.~M.~Polyakov,
``Gauge theory correlators from noncritical string theory,''
Phys. Lett. B \textbf{428} (1998), 105-114
doi:10.1016/S0370-2693(98)00377-3
[arXiv:hep-th/9802109 [hep-th]].

\bibitem{Jackiw:1984je}
R.~Jackiw,
``Lower Dimensional Gravity,''
Nucl. Phys. B \textbf{252} (1985), 343-356
doi:10.1016/0550-3213(85)90448-1
\bibitem{Teitelboim:1983ux}
C.~Teitelboim,
``Gravitation and Hamiltonian Structure in Two Space-Time Dimensions,''
Phys. Lett. B \textbf{126} (1983), 41-45
doi:10.1016/0370-2693(83)90012-6

\bibitem{Sachdev:1992fk}
S.~Sachdev and J.~Ye,
``Gapless spin fluid ground state in a random, quantum Heisenberg magnet,''
Phys. Rev. Lett. \textbf{70}, 3339 (1993)
doi:10.1103/PhysRevLett.70.3339
[arXiv:cond-mat/9212030 [cond-mat]].

\bibitem{ktt}
A.Kitaev.2015. A simple model of quantum holography, talk given at KITP strings seminar and Entanglement program, February 12, April 7, and May 27, Santa Barbara, U.S.A.

\bibitem{Maldacena:2016hyu}
J.~Maldacena and D.~Stanford,
``Remarks on the Sachdev-Ye-Kitaev model,''
Phys. Rev. D \textbf{94}, no.10, 106002 (2016)
doi:10.1103/PhysRevD.94.106002
[arXiv:1604.07818 [hep-th]].

\bibitem{Sarosi:2017ykf}
G.~S\'arosi,
``AdS$_{2}$ holography and the SYK model,''
PoS \textbf{Modave2017} (2018), 001
doi:10.22323/1.323.0001
[arXiv:1711.08482 [hep-th]].

\bibitem{Polchinski:2016xgd}
J.~Polchinski and V.~Rosenhaus,
``The Spectrum in the Sachdev-Ye-Kitaev Model,''
JHEP \textbf{04}, 001 (2016)
doi:10.1007/JHEP04(2016)001
[arXiv:1601.06768 [hep-th]].

\bibitem{Kitaev:2017awl}
A.~Kitaev and S.~J.~Suh,
``The soft mode in the Sachdev-Ye-Kitaev model and its gravity dual,''
JHEP \textbf{05}, 183 (2018)
doi:10.1007/JHEP05(2018)183
[arXiv:1711.08467 [hep-th]].

\bibitem{Jensen:2016pah}
K.~Jensen,
``Chaos in AdS$_2$ Holography,''
Phys. Rev. Lett. \textbf{117}, no.11, 111601 (2016)
doi:10.1103/PhysRevLett.117.111601
[arXiv:1605.06098 [hep-th]].


\bibitem{Das:2017eiw}
S.~R.~Das, A.~Ghosh, A.~Jevicki and K.~Suzuki,
``Duality in the Sachdev-Ye-Kitaev Model,''
Springer Proc. Math. Stat. \textbf{255}, 43-61 (2017)
doi:10.1007/978-981-13-2179-5\_4

\bibitem{Das:2017pif}
S.~R.~Das, A.~Jevicki and K.~Suzuki,
``Three Dimensional View of the SYK/AdS Duality,''
JHEP \textbf{09}, 017 (2017)
doi:10.1007/JHEP09(2017)017
[arXiv:1704.07208 [hep-th]].

\bibitem{Das:2017hrt}
S.~R.~Das, A.~Ghosh, A.~Jevicki and K.~Suzuki,
``Three Dimensional View of Arbitrary $q$ SYK models,''
JHEP \textbf{02}, 162 (2018)
doi:10.1007/JHEP02(2018)162
[arXiv:1711.09839 [hep-th]].

\bibitem{Taylor:2017dly}
M.~Taylor,
``Generalized conformal structure, dilaton gravity and SYK,''
JHEP \textbf{01}, 010 (2018)
doi:10.1007/JHEP01(2018)010
[arXiv:1706.07812 [hep-th]].


\bibitem{Jevicki:2016bwu}
A.~Jevicki, K.~Suzuki and J.~Yoon,
``Bi-Local Holography in the SYK Model,''
JHEP \textbf{07}, 007 (2016)
doi:10.1007/JHEP07(2016)007
[arXiv:1603.06246 [hep-th]].

\bibitem{Jevicki:2016ito}
A.~Jevicki and K.~Suzuki,
``Bi-Local Holography in the SYK Model: Perturbations,''
JHEP \textbf{11}, 046 (2016)
doi:10.1007/JHEP11(2016)046
[arXiv:1608.07567 [hep-th]].

\bibitem{Lala:2018yib}
A.~Lala and D.~Roychowdhury,
``SYK/AdS duality with Yang-Baxter deformations,''
JHEP \textbf{12}, 073 (2018)
doi:10.1007/JHEP12(2018)073
[arXiv:1808.08380 [hep-th]].

\bibitem{Roychowdhury:2018clp}
D.~Roychowdhury, ``Holographic derivation of $ q $ SYK spectrum with Yang-Baxter shift,'' Phys. Lett. B \textbf{797}, 134818 (2019) doi:10.1016/j.physletb.2019.134818 [arXiv:1810.09404 [hep-th]].

\bibitem{Almheiri:2014cka}
A.~Almheiri and J.~Polchinski,
``Models of AdS$_{2}$ backreaction and holography,''
JHEP \textbf{11}, 014 (2015)
doi:10.1007/JHEP11(2015)014
[arXiv:1402.6334 [hep-th]].

\bibitem{Engelsoy:2016xyb}
J.~Engels\"oy, T.~G.~Mertens and H.~Verlinde,
``An investigation of AdS$_{2}$ backreaction and holography,''
JHEP \textbf{07} (2016), 139
doi:10.1007/JHEP07(2016)139
[arXiv:1606.03438 [hep-th]].

\bibitem{Kyono:2017jtc}
H.~Kyono, S.~Okumura and K.~Yoshida,
``Deformations of the Almheiri-Polchinski model,''
JHEP \textbf{03}, 173 (2017)
doi:10.1007/JHEP03(2017)173
[arXiv:1701.06340 [hep-th]].

\bibitem{Maldacena:2016upp}
J.~Maldacena, D.~Stanford and Z.~Yang,
``Conformal symmetry and its breaking in two dimensional Nearly Anti-de-Sitter space,''
PTEP \textbf{2016}, no.12, 12C104 (2016)
doi:10.1093/ptep/ptw124
[arXiv:1606.01857 [hep-th]].


\bibitem{Gross:2017hcz}
D.~J.~Gross and V.~Rosenhaus,
``The Bulk Dual of SYK: Cubic Couplings,''
JHEP \textbf{05}, 092 (2017)
doi:10.1007/JHEP05(2017)092
[arXiv:1702.08016 [hep-th]].

\bibitem{Gross:2016kjj}
D.~J.~Gross and V.~Rosenhaus,
``A Generalization of Sachdev-Ye-Kitaev,''
JHEP \textbf{02}, 093 (2017)
doi:10.1007/JHEP02(2017)093
[arXiv:1610.01569 [hep-th]].


\bibitem{Mertens:2022irh}
T.~G.~Mertens and G.~J.~Turiaci,
``Solvable models of quantum black holes: a review on Jackiw\textendash{}Teitelboim gravity,''
Living Rev. Rel. \textbf{26} (2023) no.1, 4
doi:10.1007/s41114-023-00046-1
[arXiv:2210.10846 [hep-th]].

\bibitem{Blommaert:2023opb}
A.~Blommaert, T.~G.~Mertens and S.~Yao,
``Dynamical actions and q-representation theory for double-scaled SYK,''
JHEP \textbf{02} (2024), 067
doi:10.1007/JHEP02(2024)067
[arXiv:2306.00941 [hep-th]].
\bibitem{Lin:2022rbf}
H.~W.~Lin,
``The bulk Hilbert space of double scaled SYK,''
JHEP \textbf{11} (2022), 060
doi:10.1007/JHEP11(2022)060
[arXiv:2208.07032 [hep-th]].
\bibitem{Berkooz:2024lgq}
M.~Berkooz and O.~Mamroud,
``A Cordial Introduction to Double Scaled SYK,''
[arXiv:2407.09396 [hep-th]].

\bibitem{Blommaert:2024ydx}
A.~Blommaert, T.~G.~Mertens and J.~Papalini,
``The dilaton gravity hologram of double-scaled SYK,''
[arXiv:2404.03535 [hep-th]].

\bibitem{Blommaert:2024whf}
A.~Blommaert, A.~Levine, T.~G.~Mertens, J.~Papalini and K.~Parmentier,
``An entropic puzzle in periodic dilaton gravity and DSSYK,''
[arXiv:2411.16922 [hep-th]].

\bibitem{Brown:1986nw}
J.~D.~Brown and M.~Henneaux,
``Central Charges in the Canonical Realization of Asymptotic Symmetries: An Example from Three-Dimensional Gravity,''
Commun. Math. Phys. \textbf{104} (1986), 207-226
doi:10.1007/BF01211590

\bibitem{Castro:2008ms}
A.~Castro, D.~Grumiller, F.~Larsen and R.~McNees,
``Holographic Description of AdS(2) Black Holes,''
JHEP \textbf{11} (2008), 052
doi:10.1088/1126-6708/2008/11/052
[arXiv:0809.4264 [hep-th]].

 \bibitem{Rathi:2023vhw}
H.~Rathi and D.~Roychowdhury,
``AdS$_{2}$ holography and ModMax,''
JHEP \textbf{07} (2023), 026
doi:10.1007/JHEP07(2023)026
[arXiv:2303.14379 [hep-th]].
\bibitem{Rathi:2021aaw}
H.~Rathi and D.~Roychowdhury,
``Holographic JT gravity with quartic couplings,''
JHEP \textbf{10} (2021), 209
doi:10.1007/JHEP10(2021)209
[arXiv:2107.11632 [hep-th]].

\bibitem{Rathi:2024qsy}
H.~Rathi,
``AdS\_2/CFT\_1 at finite density and holographic aspects of 2D black holes,''
[arXiv:2404.02724 [hep-th]]

\bibitem{Kiritsis:2019npv}
E.~Kiritsis,
``String theory in a nutshell,''
Princeton University Press, 2019,
ISBN 978-0-691-15579-1, 978-0-691-18896-6

\bibitem{Calabrese:2004eu}
P.~Calabrese and J.~L.~Cardy,
``Entanglement entropy and quantum field theory,''
J. Stat. Mech. \textbf{0406} (2004), P06002
doi:10.1088/1742-5468/2004/06/P06002
[arXiv:hep-th/0405152 [hep-th]].

\bibitem{Calabrese:2009qy}
P.~Calabrese and J.~Cardy,
``Entanglement entropy and conformal field theory,''
J. Phys. A \textbf{42} (2009), 504005
doi:10.1088/1751-8113/42/50/504005
[arXiv:0905.4013 [cond-mat.stat-mech]].

\bibitem{Mertens:2017mtv}
T.~G.~Mertens, G.~J.~Turiaci and H.~L.~Verlinde,
``Solving the Schwarzian via the Conformal Bootstrap,''
JHEP \textbf{08} (2017), 136
doi:10.1007/JHEP08(2017)136
[arXiv:1705.08408 [hep-th]].

\bibitem{Murugan:2017eto}
J.~Murugan, D.~Stanford and E.~Witten,
``More on Supersymmetric and 2d Analogs of the SYK Model,''
JHEP \textbf{08} (2017), 146
doi:10.1007/JHEP08(2017)146
[arXiv:1706.05362 [hep-th]].

\bibitem{Mertens:2018fds}
T.~G.~Mertens,
``The Schwarzian theory \textemdash{} origins,''
JHEP \textbf{05} (2018), 036
doi:10.1007/JHEP05(2018)036
[arXiv:1801.09605 [hep-th]].

\bibitem{Laflorencie:2015eck}
N.~Laflorencie,
``Quantum entanglement in condensed matter systems,''
Phys. Rept. \textbf{646} (2016), 1-59
doi:10.1016/j.physrep.2016.06.008
[arXiv:1512.03388 [cond-mat.str-el]].


\bibitem{fefferman}
Fefferman, Charles; Graham, C. Robin. Conformal invariants, dans Élie Cartan et les mathématiques d'aujourd'hui - Lyon, 25-29 juin 1984, Astérisque, no. S131 (1985), 22 p.


\bibitem{formula}
Milton Abramowitz, Irene A. Stegun, ``Handbook of Mathematical Functions: with Formulas, Graphs, and Mathematical Tables (Dover Books on Mathematics)".



\bibitem{Macpherson:2014eza}
N.~T.~Macpherson, C.~N\'u\~nez, L.~A.~Pando Zayas, V.~G.~J.~Rodgers and C.~A.~Whiting,
``Type IIB supergravity solutions with AdS$_{5}$ from Abelian and non-Abelian T dualities,''
JHEP \textbf{02} (2015), 040
doi:10.1007/JHEP02(2015)040
[arXiv:1410.2650 [hep-th]].


\bibitem{Bea:2015fja}
Y.~Bea, J.~D.~Edelstein, G.~Itsios, K.~S.~Kooner, C.~Nunez, D.~Schofield and J.~A.~Sierra-Garcia,
``Compactifications of the Klebanov-Witten CFT and new AdS$_{3}$ backgrounds,''
JHEP \textbf{05} (2015), 062
doi:10.1007/JHEP05(2015)062
[arXiv:1503.07527 [hep-th]].

\bibitem{Merrikin:2022yho}
P.~Merrikin, C.~Nunez and R.~Stuardo,
``Compactification of 6d N=(1,0) quivers, 4d SCFTs and their holographic dual Massive IIA backgrounds,''
Nucl. Phys. B \textbf{996} (2023), 116356
doi:10.1016/j.nuclphysb.2023.116356
[arXiv:2210.02458 [hep-th]].

\bibitem{Chatzis:2024kdu}
D.~Chatzis, A.~Fatemiabhari, C.~Nunez and P.~Weck,
``SCFT deformations via uplifted solitons,''
Nucl. Phys. B \textbf{1006} (2024), 116659
doi:10.1016/j.nuclphysb.2024.116659
[arXiv:2406.01685 [hep-th]].
\bibitem{Garcia-Garcia:2020ttf}
A.~M.~Garc\'\i{}a-Garc\'\i{}a and V.~Godet,
``Euclidean wormhole in the Sachdev-Ye-Kitaev model,''
Phys. Rev. D \textbf{103} (2021) no.4, 046014
doi:10.1103/PhysRevD.103.046014
[arXiv:2010.11633 [hep-th]].
\bibitem{Rathi:2021mla}
H.~Rathi and D.~Roychowdhury,
``Phases of Euclidean wormholes in JT gravity,''
Nucl. Phys. B \textbf{994} (2023), 116315
doi:10.1016/j.nuclphysb.2023.116315
[arXiv:2111.11279 [hep-th]].

\bibitem{Blommaert:2025avl}
A.~Blommaert, A.~Levine, T.~G.~Mertens, J.~Papalini and K.~Parmentier,
``Wormholes, branes and finite matrices in sine dilaton gravity,''
[arXiv:2501.17091 [hep-th]].

\bibitem{Maldacena:2019cbz}
J.~Maldacena, G.~J.~Turiaci and Z.~Yang,
``Two dimensional Nearly de Sitter gravity,''
JHEP \textbf{01} (2021), 139
doi:10.1007/JHEP01(2021)139
[arXiv:1904.01911 [hep-th]].
\bibitem{Alonso-Monsalve:2024oii}
E.~Alonso-Monsalve, D.~Harlow and P.~Jefferson,
``Phase space of Jackiw-Teitelboim gravity with positive cosmological constant,''
[arXiv:2409.12943 [hep-th]].
\bibitem{Held:2024rmg}
J.~Held and H.~Maxfield,
``The Hilbert space of de Sitter JT: a case study for canonical methods in quantum gravity,''
[arXiv:2410.14824 [hep-th]].

\bibitem{Rahman:2022jsf}
A.~A.~Rahman,
``dS JT Gravity and Double-Scaled SYK,''
[arXiv:2209.09997 [hep-th]].
\bibitem{Susskind:2022bia}
L.~Susskind,
``De Sitter Space, Double-Scaled SYK, and the Separation of Scales in the Semiclassical Limit,''
[arXiv:2209.09999 [hep-th]].

\bibitem{Verlinde:2024zrh}
H.~Verlinde and M.~Zhang,
``SYK Correlators from 2D Liouville-de Sitter Gravity,''
[arXiv:2402.02584 [hep-th]].


\end{thebibliography}
\end{document}